# Description of Even-Even $^{114-134}Xe$ Isotopes in the Transitional Region of IBM

M. A. Jafarizadeh[a,b1], N. Fouladi[c], H. Sabri[c2]

[a]Department of Theoretical Physics and Astrophysics, University of Tabriz, Tabriz 51664, Iran.

[b]Research Institute for Fundamental Sciences, Tabriz 51664, Iran.

[c]Department of Nuclear Physics, University of Tabriz, Tabriz 51664, Iran.

---

[1] E-mail: jafarizadeh@tabrizu.ac.ir
[2] E-mail: h-sabri@tabrizu.ac.ir

## Abstract

Properties of $^{114-134}Xe$ isotopes are studied in the $U(5) \leftrightarrow SO(6)$ transitional region of Interacting Boson Model (IBM-1). The energy levels and $B(E2)$ transition rates are calculated via the affine $SU(1,1)$ Lie Algebra. The agreement with the most recent experimental is acceptable. The evaluated Hamiltonian control parameters suggest a spherical to $\gamma-$soft shape transition and propose the $^{130}Xe$ nucleus as the best candidate for the $E(5)$ symmetry.

## Introduction

General algebraic group techniques, applied to the Interacting Boson Model (IBM), have rather successfully described the low-lying collective properties of a wide range of nuclei. In the relatively simple Hamiltonian of the model, the collective states are described by a system of interacting $s$- and $d$-bosons carrying angular momenta 0 and 2, respectively, which define an overall $U(6)$ symmetry [1-3]. The IBM Hamiltonian has exact solutions in three dynamical symmetry limits [$U(5), O(6)$ and $SU(3)$], which are geometrically analogous to the anharmonic vibrator, axial rotor and $\gamma-$unstable rotor, respectively. More generally, the Hamiltonian can be expressed in terms of an invariant operator of that chain of symmetries, and a shape phase transition between the dynamical symmetry limits results [4-6]. The analytic description of the structural change at the critical point of the phase transition being still an open problem, the Hamiltonian must be diagonalized numerically. Pan *et al* [7], proposed a new solution based on the affine $SU(1,1)$ algebraic technique, which determines the properties of nuclei in the $U(5) \leftrightarrow SO(6)$ transitional region of IBM-1[7-8].

The xenon isotopes have been previously analyzed both theoretically and experimentally [9-27] with particular emphasis on describing the experimental data via collective models. The ground state properties of even–even $Xe$ isotopes have been the subject to theoretical [9-18] and experimental studies [20] involving in-beam γ -ray spectroscopy. Recently, the nuclear structure of xenon isotopes have been investigated by Turkan in the IBM-1 model [28], while the energy levels, electric quadrupole moments and $B(E2)$ values of even-mass nuclei such as $Ba$, $Xe$ were studied within the framework of the IBM-2 [29-33]. These descriptions suggest these nuclei to be soft with regard to $\gamma$ deformation with a nearly maximum effective trixiality of $\gamma \cong 30°$ [9]. As pointed out by Zamfir et al [9], $Xe$ isotopes in the mass region $A \sim 130$ appear to evolve from $U(5)-$ to $O(6)$-like structure in the IBM-1. It is very difficult to apply conventional mean-field theories to such structures, which are neither vibrational nor rotational.

Here we examine the even- even $^{114-134}Xe$ isotopes in the $U(5) \leftrightarrow SO(6)$ transition region and calculate the energy levels and $B(E2)$ transition probabilities in the frame work of IBM with the affine $SU(1,1)$ algebraic technique. The estimated control parameters indicate a spherical to $\gamma-$ soft shape transition. The same shape transition is revealed by the evolution of two-neutron separation energies $S_{2n}$ [29] derived from experimental results [30-33]. Also, special values are found for the $^{130}Xe$ control parameter and $R_{4/2}$, which suggest it as the best candidate for $E(5)$ dynamical symmetry in this isotopic chain.

This paper is organized as follows: section 2 briefly summarizes the theoretical aspects of transitional Hamiltonian and the affine $SU(1,1)$ algebraic technique. Section 3 presents the numerical results obtained from applying the considered Hamiltonian to different isotopes. Finally, Section 4 summarizes our findings and the conclusions extracted from the results in section 3.

## 2. Theoretical framework

The phenomenological Interacting Boson Model (IBM) in terms of $U(5)$, $O(6)$ and $SU(3)$ dynamical symmetries has been employed to describe the collective properties of several medium- and heavy-mass nuclei. These dynamical symmetries are geometrically analogues to the harmonic vibrator, axial rotor and $\gamma-$unstable rotor, respectively [1-3]. While these symmetries have already offered a fairly accurate description of the low-lying nuclear states, attempts to analytically describe the structure at the critical point of the phase transition have only been partially successful. Iachello [4-5] established a new set of dynamical symmetries, i.e. $E(5)$ and $X(5)$, for nuclei located at critical point of transitional regions. The $E(5)$ symmetry, which describes a second order phase transition, corresponds to the transitional states in the region from the $U(5)$ to the $O(6)$ symmetries in the IBM-1. Different analyses of this transitional region suggested certain nuclei, such as $^{134}Ba$, $^{108}Pd$, as examples of that symmetry [7.12].

Elaborate numerical techniques are required to diagonalize the Hamiltonian in these transitional regions and critical points. To avoid these problems, an algebraic solution based on the affine $SU(1,1)$ Lie algebra has been proposed by Pan et al [7-8] to describe the properties of nuclei located in the $U(5) \leftrightarrow SO(6)$ transitional region. The results of this approach are some what different from those obtained from the IBM, but as pointed out in Refs.[7-8], there is a clear correspondence with the description of the geometrical model for this transitional region.

## 2.1. The affine su(1,1) approach to the transitional Hamiltonian

References 7 and 8 describe the $SU(1,1)$ Algebra in detail. Here, we briefly outline the basic ansatz and summarize the results. The Lie algebra corresponding to the group $SU(1,1)$ is generated by the operators $S^v, v = 0$ and $\pm$, which satisfies the following commutation relations

$$[S^0, S^\pm] = \pm S^\pm \quad , \quad [S^+, S^-] = -2S^0 \tag{2.1}$$

The Casimir operator of $SU(1,1)$ can be written as

$$\hat{C}_2 = S^0(S^0 - 1) - S^+ S^- \tag{2.2}$$

Representations of $SU(1,1)$ are determined by a single number $\kappa$. The representation of the Hilbert space is hence spanned by orthonormal basis $|\kappa\mu\rangle$, where $\kappa$ can be any positive number and $\mu = \kappa, \kappa+1, \dots$. Therefore,

$$\hat{C}_2(SU(1,1))|\kappa\mu\rangle = \kappa(\kappa - 1)|\kappa\mu\rangle \quad , \quad S^0|\kappa\mu\rangle = \mu|\kappa\mu\rangle \tag{2.3}$$

The bases of $U(5) \supset SO(5)$ and $SO(6) \supset SO(5)$ are simultaneously the bases of $SU^d(1,1) \supset U(1)$ and $SU^{sd}(1,1) \supset U(1)$, respectively. For $U(5) \supset SO(5)$ case, one has

$$|N n_d \nu n_\Delta L M\rangle = \left|N, \kappa^d = \frac{1}{2}(\nu + \frac{5}{2}), \mu^d = \frac{1}{2}(n_d + \frac{5}{2}), n_\Delta L M\right\rangle \tag{2.4}$$

where $N, n_d, \nu, L$ and $M$ are quantum numbers of $U(6), U(5), SO(5), SO(3)$ and $SO(2)$, respectively, while $n_\Delta$ is an additional quantum number needed in the reduction $SO(5) \downarrow SO(3)$ and $\kappa^d$ and $\mu^d$ are quantum numbers of $SU^d(1,1)$ and $U(1)$ respectively. On the other hand, in IBM-1, the generators of the $d-$boson pairing algebra created by

$$S^+(d) = \frac{1}{2}(d^\dagger \cdot d^\dagger) \quad , \quad S^-(d) = \frac{1}{2}(\tilde{d} \cdot \tilde{d}) \quad , \quad S^0(d) = \frac{1}{4}\sum_\nu (d^\dagger_\nu d_\nu + d_\nu d^\dagger_\nu) \tag{2.5}$$

Similarly, the $s-$boson pairing forms another $SU^s(1,1)$ algebra generated by the operators

$$S^+(s) = \frac{1}{2}s^{\dagger 2} \quad , \quad S^-(s) = \frac{1}{2}s^2 \quad , \quad S^0(s) = \frac{1}{4}(s^\dagger s + s s^\dagger) \tag{2.6}$$

The infinite dimensional $SU(1,1)$ algebra is then generated by the operators [7-8]

$$S^\pm_n = c_s^{2n+1} S^\pm(s) + c_d^{2n+1} S^\pm(d) \quad , \quad S^0_n = c_s^{2n} S^0(s) + c_d^{2n} S^0(d) \tag{2.7}$$

Where $c_s$ and $c_d$ are real parameters and $n$ can be $0, \pm 1, \pm 2, \dots$.
The generators in Eq.(2.7) satisfy the commutation relations

$$[S^0_m, S^\pm_n] = \pm S^\pm_{m+n} \quad , \quad [S^+_m, S^-_n] = -2S^0_{m+n+1} \tag{2.8}$$

It follows that the $\{S^\mu_m, \mu = 0, +, -; \pm 1, \pm 2, \dots\}$ generate an affine Lie algebra $\widehat{SU(1,1)}$ without central extension. From the generators of the Algebra, the following Hamiltonian for transitional region between $U(5) \leftrightarrow SO(6)$ limits can then be written [7-8]

$$\hat{H} = g\, S_0^+ S_0^- + \varepsilon\, S_1^0 + \gamma\, \hat{C}_2(SO(5)) + \delta\, \hat{C}_2(SO(3)) \tag{2.9}$$

$g, \varepsilon, \gamma$ and $\delta$ are real parameters and $\hat{C}_2(SO(3))$ and $\hat{C}_2(SO(5))$ denote the Casimir operators of these groups. It can be seen that the Hamiltonian (2.9), would be equivalent with the $SO(6)$ Hamiltonian if $c_s = c_d$ and to the $U(5)$ Hamiltonian if $c_s = 0$ & $c_d \neq 0$. Therefore, the inequalities $c_s \neq c_d \neq 0$ correspond to the $U(5) \leftrightarrow SO(6)$ transitional region. In our calculation, we let $c_d$ be a constant(=1) and $c_s$ vary between 0 and $c_d$.

In order to obtain the eigenstates of Hamiltonian (2.9), with exploit a Fourier-Laurent expansion of the eigenstates and the generators in terms of unknown $c-$number parameters $x_i$ $(i=1,2,...,k)$. in other words, we write the eigenstates in the form [7-8]

$$|k; v_s v n_\Delta L M\rangle = \sum_{n_i \in Z} a_{n_1} a_{n_2} ... a_{n_k} x_1^{n_1} x_2^{n_2} ... x_k^{n_k} S_{n_1}^+ S_{n_2}^+ ... S_{n_k}^+ |lw\rangle \tag{2.10}$$

Given the analytical behavior of the wavefunctions, it suffices to consider $x_i$ near zero. The commutation relations (2.3) between the generators of $SU(1,1)$ Algebra (2.3), the allow us to express the wavefunctions as:

$$|k; v_s v n_\Delta L M\rangle = N S_{x_1}^+ S_{x_2}^+ ... S_{x_k}^+ |lw\rangle \tag{2.11}$$

where $N$ is a normalization factor and

$$S_{x_i}^+ = \frac{c_s}{1 - c_s^2 x_i} S^+(s) + \frac{c_d}{1 - c_d^2 x_i} S^+(d) \tag{2.12}$$

The c-numbers $x_i$ are determined by the following set of equations ($v_s$ denotes the quantum number of $SO(5)$ group for $s$ bosons)

$$\frac{\epsilon}{x_i} = \frac{g c_s^2 (v_s + \frac{1}{2})}{1 - c_s^2 x_i} + \frac{g c_d^2 (v + \frac{5}{2})}{1 - c_d^2 x_i} - \sum_{i \neq j} \frac{2}{x_i - x_j} \qquad \text{for } i=1,2,...,k \tag{2.13}$$

The eigenvalues $E^{(k)}$ of Hamiltonian (2.9) can be expressed in the form [7-8]

$$E^{(k)} = h^{(k)} + \gamma v(v+3) + \delta L(L+1) + \varepsilon \Lambda_1^0 \qquad , \qquad \Lambda_1^0 = \frac{1}{2}[c_s^2(v_s + \frac{1}{2}) + c_d^2(v + \frac{5}{2})] \tag{2.14}$$

Where

$$h^{(k)} = \sum_{i=1}^{k} \frac{\varepsilon}{x_i} \tag{2.15}$$

The quantum number $k$ is related to total boson number by the equality

$$N = 2k + v_s + v$$

To obtain numerical results for $E^{(k)}$ (energy spectra of considered nuclei), we have followed the prescriptions introduced in Refs.7 and 8, i.e., we solve a set of non-linear Bethe-Ansatz equations with $k$ unknowns for $k$ pair excitations. must be solved. It is convenient to change variables as follows

$$\epsilon = \frac{\varepsilon}{g} (g=1\ kev\ [7\text{-}8]) \qquad c = \frac{c_s}{c_d} \leq 1 \qquad y_i = c_d^2 x_i$$

To rewrite Eq.(2.13) in the form

$$\frac{\epsilon}{y_i} = \frac{c^2(\nu_s + \frac{1}{2})}{1 - c^2 y_i} + \frac{(\nu + \frac{5}{2})}{1 - y_i} - \sum_{i \neq j} \frac{2}{y_i - y_j} \qquad \text{for i=1,2,...,k} \qquad (2.16)$$

To determine the roots of Bethe-Ansatz equations with specified values of $\nu_s$ and $\nu$, we solve Eq. (2.16) with definite values of $c$ and $\varepsilon$ for $i = 1$ and then use the function "Find root" in Maple13 to obtain all $y'_j$ s. We then repeat this procedure with different $c$ and $\varepsilon$ to minimize the deviation $\sigma$ between the energy spectra (after inserting $\gamma$ and $\delta$) and the experimental values. The deviation is defined by the equality

$$\sigma = \left( \frac{1}{N_{tot}} \sum_{i, tot} \left| E_{\exp}(i) - E_{cal}(i) \right|^2 \right)^{1/2}$$

Where $N_{tot}$ is the number of energy levels in the fit. To optimize the set of Hamiltonian parameters $\gamma$ and $\delta$, we have carried out a least-square fit to the available experimental data [36-39] of the excitation energies for selected states, $0_1^+, 2_1^+, 4_1^+, 0_2^+, 2_2^+, 4_2^+$, etc (12 levels up to $2_4^+$, although not all of them are available for all considered nuclei) or of the two neutron separation energies of considered nuclei.

## 2.2. $B(E2)$ Transition

Additional information on the structure of nuclei can be obtained from other observables: the reduced electric quadrupole transition probabilities $B(E2)$ and quadrupole moment ratios within the low-lying. The E2 transition operator must be a Hermitian tensor of rank two; consequently, the number of bosons must be conserved. These constraints limit to two the number of allowed in lowest order, the electric quadrupole transition operator being given by the expression [7],

$$\hat{T}_\mu^{(E2)} = q_2 [\hat{d}^\dagger \times \tilde{s} + \hat{s}^\dagger \times \tilde{d}]_\mu^{(2)} + q'_2 [\hat{d}^\dagger \times \tilde{d}]_\mu^{(2)} \qquad , \qquad (2.17)$$

Where $q_2$ is the effective quadrupole charge, $q'_2$ is a dimensionless coefficient, and $s^\dagger (d^\dagger)$ is the creation operator of $s(d)$ boson. The reduced electric quadrupole transition rates between $I_i \rightarrow I_f$ states are given by [3]

$$B(E2; I_i \rightarrow I_f) = \frac{\left| \langle I_f \| T(E2) \| I_i \rangle \right|}{2I_i + 1} \qquad , \qquad (2.18)$$

To determine the $q_2$ and $q'_2$, we have followed the procedure in Refs. 7 and 8, i.e. treated these parameters as function of total boson number $N$.

## 3. Numerical result

### 3.1. Energy levels

The experimental energy spectra [9-20], suggest that we collect empirical evidence concerning the $U(5) \leftrightarrow SO(6)$ transitional region from $^{114-134}Xe$ isotopes. We have, therefore computed the energy spectra for the transition-region Hamiltonian of the (2.9). Figure 1 displays 12 levels, up to $2_4^+$, for an illustrative set of nuclei in our fitting procedure. Table 1 shows the optimal Hamiltonian parameters $\varepsilon, c_s, \delta$ and $\gamma$ resulting from the procedure in Section 2, i.e., the parameters minimizing the deviation $\sigma$ calculated from the experimental data in Refs. 36-39. Shown in Fig.1 are the available experimental levels and corresponding calculated levels for $^{120}Xe - ^{126}Xe$ isotopes in the low-lying region of the spectra. The agreement is acceptable.

| Nucleus | N | $\varepsilon(kev)$ | $c_s$ | $\gamma(kev)$ | $\delta(kev)$ | $\sigma$ |
|---|---|---|---|---|---|---|
| $^{114}_{54}Xe$ | 7 | 800 | 0.66 | 54.83 | −60.01 | 151 |
| $^{116}_{54}Xe$ | 8 | 1430 | 0.59 | 21.44 | −56.87 | 112 |
| $^{118}_{54}Xe$ | 9 | 755 | 0.79 | −52.87 | 39.10 | 89 |
| $^{120}_{54}Xe$ | 10 | 620 | 0.86 | −45.96 | 41.24 | 104 |
| $^{122}_{54}Xe$ | 9 | 540 | 0.95 | −29.64 | 30.90 | 68 |
| $^{124}_{54}Xe$ | 8 | 680 | 0.89 | −61.36 | 43.36 | 75 |
| $^{126}_{54}Xe$ | 7 | 695 | 0.83 | −56.05 | 43.77 | 91 |
| $^{128}_{54}Xe$ | 6 | 1570 | 0.65 | −138.83 | 43.11 | 88 |
| $^{130}_{54}Xe$ | 5 | 1100 | 0.46 | −79.58 | 39.71 | 105 |
| $^{132}_{54}Xe$ | 4 | 1680 | 0.37 | −114.77 | 40.31 | 115 |
| $^{134}_{54}Xe$ | 3 | 670 | 0.06 | 1.85 | 20.14 | 73 |

Table1.The parameters of the Hamiltonian (2.9) determined by least-square fitting to the experimental data for different $Xe$ isotopes. $N$ is the boson number and $\varepsilon, c_s, \gamma$ and $\delta$ are the parameters of transitional Hamiltonian (2.9) for each nuclei. The deviation $\sigma$ monitors the quality of the fitting.

## 3.2. Transition probabilities

The stable even-even nuclei in $Xe$ isotopic chain offer an excellent opportunity to study the behavior of the total low-lying E2 strengths in the transitional region from deformed to spherical nuclei. The computation of the electromagnetic transition probabilities provides a reliable test of the nuclear-model wave functions. To determine the boson effective charges $q_2$ and $q'_2$, we fit the theoretical results to the empirical $B(E2)$ values, taking the two parameters to be function of the total boson number $N$ [7-8]. The theoretical $B(E2)$ transition rates, which displayed in Figure2, are associated with the effective charge parameters in Table2.

| Nucleus | $q_2$ | $q_2'$ | Nucleus | $q_2$ | $q_2'$ |
|---|---|---|---|---|---|
| $^{114}_{54}Xe$ | 0.136 | −0.335 | $^{116}_{54}Xe$ | 0.148 | −0.415 |
| $^{118}_{54}Xe$ | 0.154 | −0.515 | $^{120}_{54}Xe$ | 0.164 | −0.637 |
| $^{122}_{54}Xe$ | 0.149 | −0.502 | $^{124}_{54}Xe$ | 0.141 | −0.398 |
| $^{126}_{54}Xe$ | 0.133 | −0.324 | $^{128}_{54}Xe$ | 0.128 | −0.271 |
| $^{130}_{54}Xe$ | 0.121 | −0.219 | $^{132}_{54}Xe$ | 0.113 | −0.176 |
| $^{134}_{54}Xe$ | 0.106 | −0.143 | | | |

Table2. Coefficients $q_2$ and $q_2'$ resulting from our analysis, similar to the procedure in Refs. 7 and 8. The corresponding $B(E2)$ values are compared to the experimental data in Figure2.

Figure 2 compares our results for $B(E2;2_1^+ \to 0_1^+)$, $B(E2;4_1^+ \to 2_1^+)$ and the ratio $B(E2;4_1^+ \to 2_1^+)/B(E2;2_1^+ \to 0_1^+)$ with experimental values [36-39]. In all figures in this paper, the experimental uncertainties are smaller than the symbols. The good agreements in Figs. 1 and 2 attest to the reliability of the fitting procedure and of our computation of the $B(E2)$ transition probabilities of even-even $Xe$ isotopes, respectively. The control parameters in Table 1 moreover provide information on the structural changes in nuclear deformation and shape phase transition.

The ground state two-neutron separation energies $S_{2n}$ are sensitive to the details of nuclear structure. Gross nuclear structure features, such as major shell closures, are clearly seen in the evolution of this observable along the isotopic chains [34-35,40]. Zamfir *et al* [35] have suggested that $S_{2n}$ vary smoothly as the nuclei undergo a second order shape phase transition between spherical ($U(5)$) and $\gamma$−unstable rotor ($SO(6)$) limits. The correlations between the two observables, one $S_{2n}$ related to ground state properties and other, $R_{4/2}$, related to the properties of the excited states, is a convenient probe of the shape phase transition region. In order to bring to light the nuclear-structure information in the two observables, we studied the evolution of the two neutron separation energies ($S_{2n}$) along the isotopic chains for the even-even $Xe$ nuclei. Experimental and theoretical values are presented in Fig. 3, including the last review of nuclear masses in Ref. 35 and the most recent available data [36-39]. On the theoretical side, to determine $S_{2n}$ in the framework of the IBM-1, we have followed the prescription in Ref.34. According to Iachello's definition, as a function of proton and neutron number, the binding energy is given by [34]

$$E_B(N_p, N_n) = E^{(c)} + A_p N_p + A_n N_n + \frac{1}{2} B_p N_p (N_p - 1) + \frac{1}{2} B_n N_n (N_n - 1) + C N_p N_n + \tilde{E}(N_p, N_n) , \quad (3.1)$$

where $N_p(N_n)$ is the number of proton(neutron) bosons in the valence shell, $E^{(c)}$ the contribution from the core and $\tilde{E}$ is the contribution to the binding energy due to the deformation. With using the Eq.(3.1), one obtains the following relation for the two neutron separation energy

$$S_{2n}(N_p, N_n) = E_B(N_p, N_n) - E_B(N_p, N_n - 1) = A_n + BN_p + C_n N_n + [\tilde{E}(N_p, N_n) - \tilde{E}(N_p, N_n - 1)], \qquad (3.2)$$

The $Xe$ isotopes have $N_p = 2$ but different numbers of neutron bosons. Letting $A_n + B = 23.70 Mev$ and $C_n = 0.814 Mev$, we obtain the two neutron separation energies compared with experimental values in Fig.3, which shows good agreement. The results confirm the predictions by Zamfir *et al.* and suggest that the phase transition for this chain of $Xe$ isotopes is of second order.

The shape phase transition is associated with a sudden change in nuclear collective behavior, as a result of which the ratio $R_{4/2} = E_{4_1^+}/E_{2_1^+}$ suddenly increase, from the spherical vibrator value of 2.0 to the deformed $\gamma$ − soft nuclei value of 2.5. Iachello proposed the value 2.20 for the $E(5)$ dynamical symmetry characterizing the critical point of $U(5) \leftrightarrow SO(6)$ transitional region [5]. Table 3 shows the estimated control parameters $c_s$ and the ratio $R_{4/2}$ for the isotopic chain. The evolution of these quantities between spherical ($c_s = 0$ & $R_{4/2} = 2.0$ for $U(5)$ limit) and $\gamma$ − unstable ($c_s = 1$ & $R_{4/2} = 2.5$ for $SO(6)$ limit) shapes, are in line with the second-order shape phase transition highlighted in our discussion of Fig. 3.

| Nuclei | $^{114}_{54}Xe$ | $^{116}_{54}Xe$ | $^{118}_{54}Xe$ | $^{120}_{54}Xe$ | $^{122}_{54}Xe$ | $^{124}_{54}Xe$ | $^{126}_{54}Xe$ | $^{128}_{54}Xe$ | $^{130}_{54}Xe$ | $^{132}_{54}Xe$ | $^{134}_{54}Xe$ |
|---|---|---|---|---|---|---|---|---|---|---|---|
| $c_s$ | 0.66 | 0.59 | 0.79 | 0.86 | 0.95 | 0.89 | 0.83 | 0.65 | 0.46 | 0.37 | 0.06 |
| $R_{4/2} = E_{4_1^+}/E_{2_1^+}$ | 2.38 | 2.33 | 2.40 | 2.47 | 2.50 | 2.48 | 2.42 | 2.33 | 2.24 | 2.15 | 2.02 |

Table3. The control parameters $c_s$ and $R_{4/2}$ ratio for the considered nuclei. The special values of the two parameters identify $^{130}Xe$ as the best candidate for $E(5)$ dynamical symmetry.

The variation of the control parameters, $c_s \sim 0 \rightarrow 1$, indicates structural changes in nuclear deformation and shape phase transitions in even-even $^{114-134}Xe$ isotopes. Iachello took $\eta$ to be the control parameter in his description of the shape phase transition [4], so that the critical points of the transitional regions are expected at or near $\eta = 0.5$. By the same token, we expect the $E(5)$ symmetry to arise at or near $c_s = 0.5$ in our approach. The control parameters and ratios in Table 3 give evidence favoring the notion of $E(5)$ symmetry for $^{130}Xe$ [41-42], which displays values of $c_s$ and $R_{4/2}$ that come closet to Iachello 's prediction, $R_{4/2} \sim 2.24$ and $c_s \sim 0.46$.

## 4. CONCLUSIONS

We have employed an affine $SU(1,1)$ Lie algebra to calculate the energies and $B(E2)$ transition probabilities for $^{114-134}Xe$ nuclei within the framework of the Interacting Boson Model. We have checked the validity of the parameters chosen in our formulation of the IBM-1 Hamiltonian and found satisfactory agreement between the presented and experimental data. Our study accounts for the general characteristics of the $Xe$ isotopes and support the notion of shape coexistence. Figures 1-3 show acceptable agreement between the presented IBM-1 results and the experimental results for the considered nuclei. The $Xe$ being close to both proton and neutron closed shells, these nuclei are not expected to be deformed. Gamma-soft rotor features exist in $Xe$ isotopes but the vibrational character is dominant. Our results confirm the adequacy of the method to describe the structure of nuclei around mass $A \sim 130$.

# Figure caption

**Figure1.** Comparison of calculated energy levels and experimental spectra taken from Refs.[36-39] for $^{120-126}Xe$ isotopes. Due to similar correspondences, we wouldn't present this comparison for other isotopes.

**Figure2.** Comparison of calculated transition probabilities and corresponding experimental values taken from Refs. [36-39]. $B(E2;2_1^+ \to 0_1^+)$ for $^{114-134}Xe$ isotopes, $B(E2;4_1^+ \to 2_1^+)$ for $^{114-134}Xe$ except ($^{126,130}Xe$). The figure also indicates the calculated $B_{4/2}$ ratios and experimental ones for $^{114-134}Xe$ except ($^{114,126,130}Xe$) nuclei.

**Figure3.** The experimental and theoretical $S_{2n}$ energies (in $kev$) for considered nuclei.

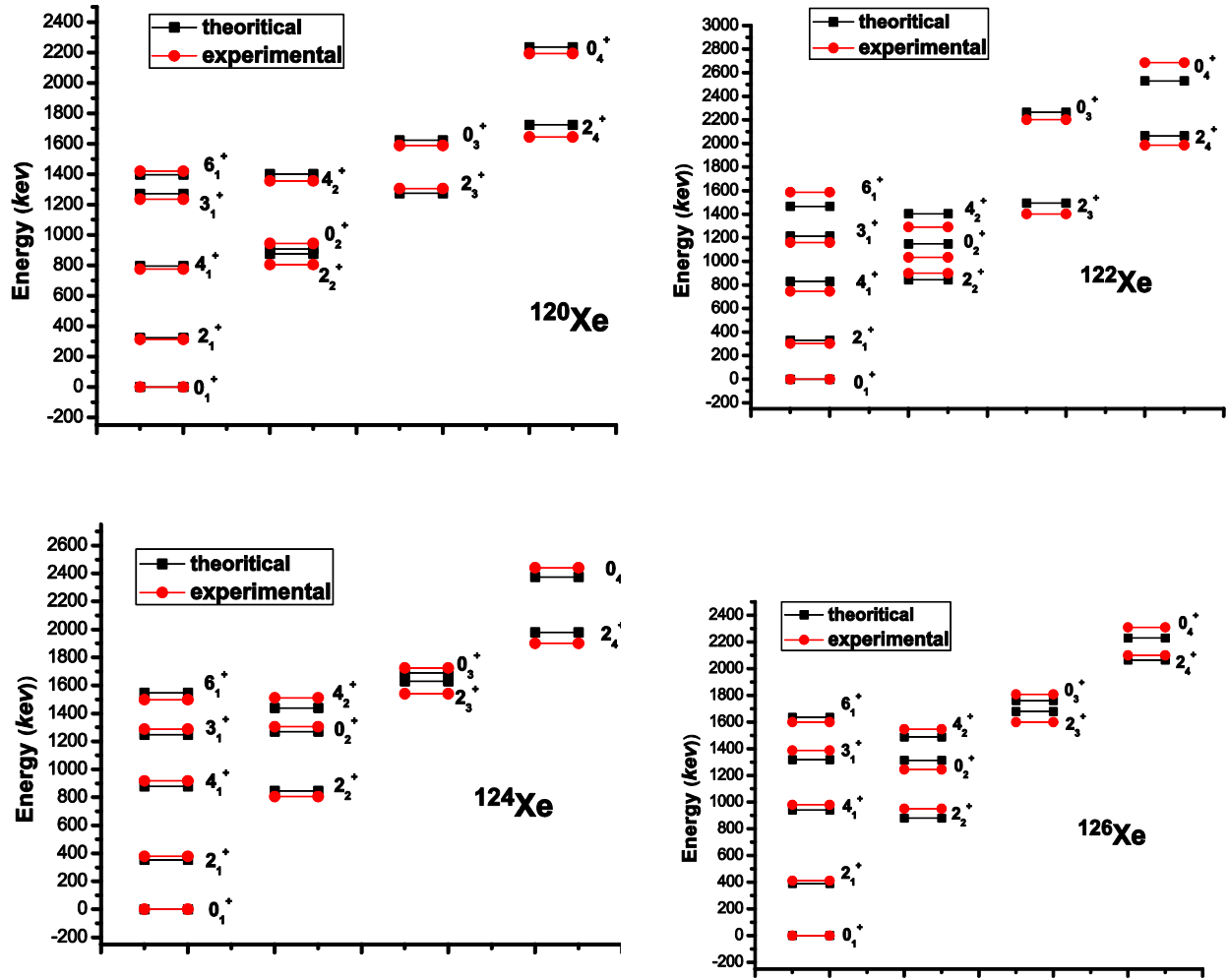

Figure2.

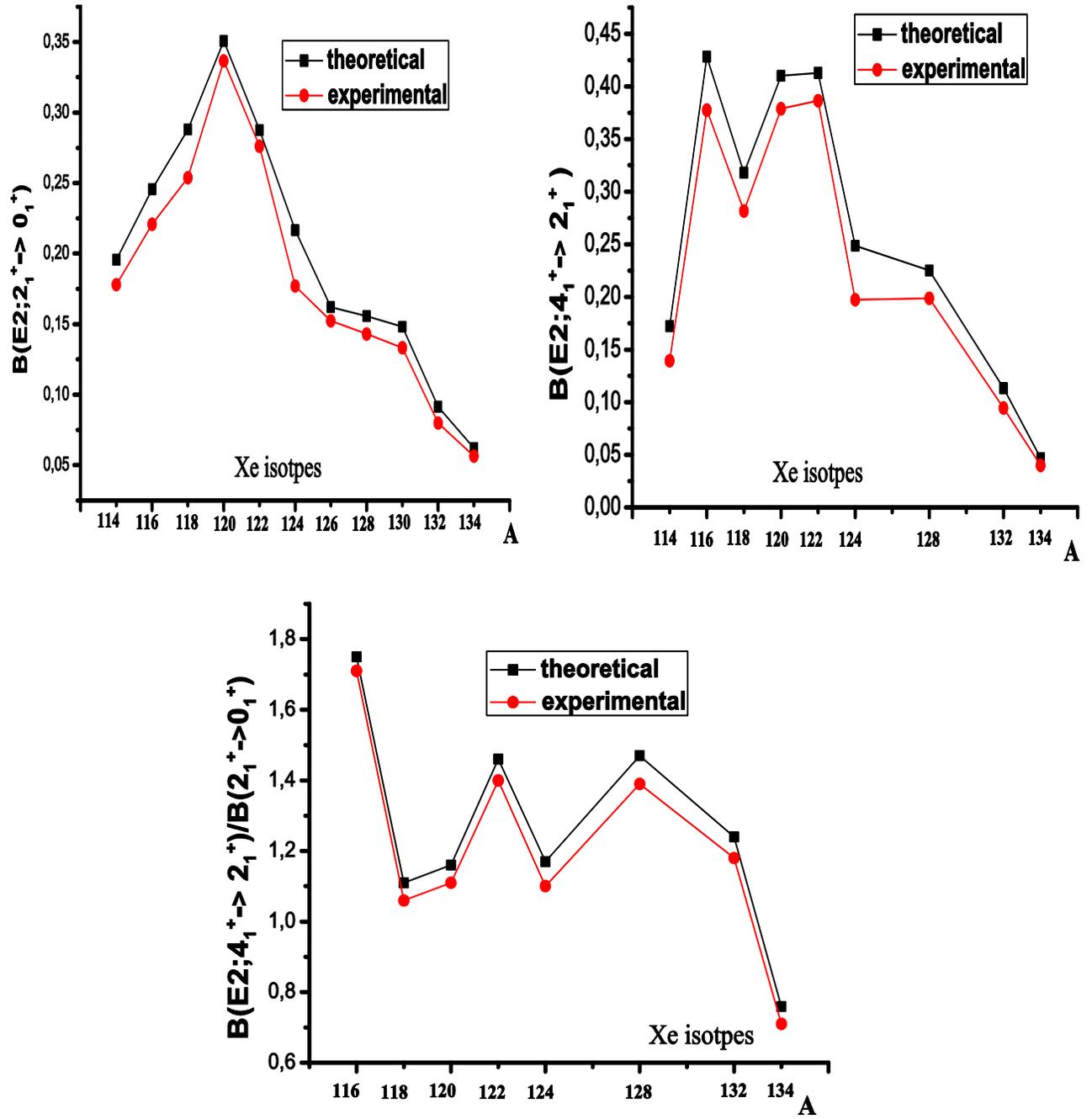

Figure3.

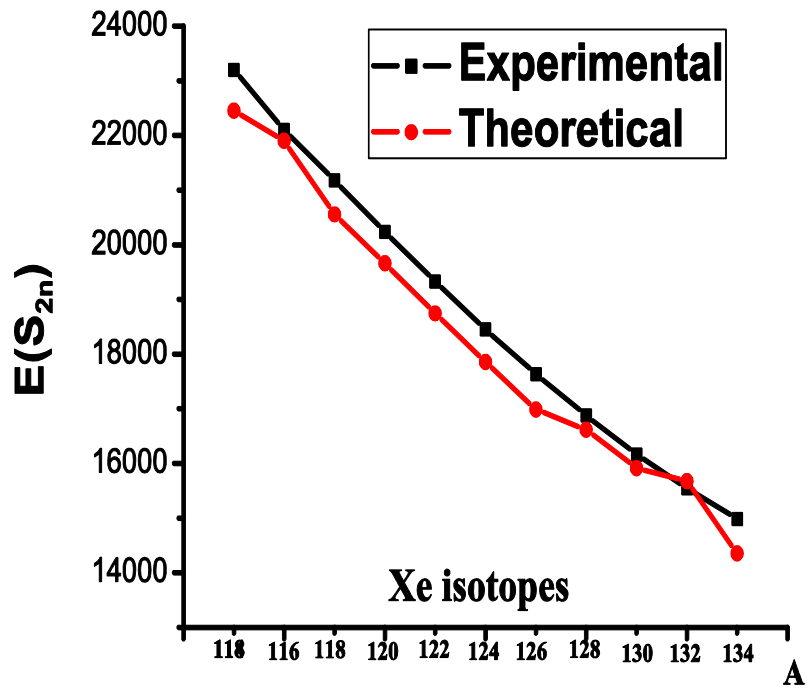